\documentclass[traditabstract]{aa}
\usepackage{txfonts}
\usepackage{natbib}
\usepackage{graphicx}
\usepackage{wrapfig}
\usepackage{color}
\usepackage{hyperref} 
\bibpunct{(}{)}{;}{a}{}{,} 

\begin{document}

\title{Synchrotron pair halo and echo emission from blazars in the cosmic web: application to extreme TeV blazars}

\author{Foteini Oikonomou\inst{1,2}
  \and Kohta Murase\inst{3}
     \and Kumiko Kotera\inst{1,4}}

\offprints{\email{ oikonomou@psu.edu }}

\institute{
Astrophysics Group, Department of Physics and Astronomy, University College London,
Gower Street, London WC1E 6BT, United Kingdom.\and
Department of Physics, Pennsylvania State University, University Park, PA 16802,USA.\and
Hubble Fellow -- Institute for Advanced Study, 1 Einstein Dr. Princeton, NJ 08540, USA.\and
Institut d'Astrophysique de Paris
UMR7095 -- CNRS, Universit\'e Pierre \& Marie Curie,
98 bis boulevard Arago
F-75014 Paris, France.
}

\authorrunning{F. Oikonomou et al.}
\titlerunning{Synchrotron pair halo and echo emission from blazars in the cosmic web}

\date{\today}

\abstract{High frequency peaked, high redshift blazars, are extreme in the sense that their spectrum is particularly hard and peaks at TeV energies. Standard leptonic scenarios require peculiar source parameters and/or a special setup in order to account for these observations. Electromagnetic cascades seeded by ultra-high energy cosmic rays (UHECR) in the intergalactic medium have also been invoked, assuming a very low intergalactic magnetic field (IGMF). Here we study the synchrotron emission of UHECR secondaries produced in blazars located in magnetised environments, and show that it can provide an alternative explanation to these challenged channels, for sources embedded in structured regions with magnetic field strengths of the order of $10^{-7}$~G. To demonstrate this, we focus on three extreme blazars: 1ES 0229+200, RGB J0710+591, and 1ES 1218+304. We model the expected gamma-ray signal from these sources through a combination of numerical Monte Carlo simulations and solving the kinetic equations of the particles in our simulations, and explore the UHECR source and intergalactic medium parameter space to test the robustness of the emission. 
We show that the generated synchrotron pair halo/echo flux at the peak energy is not sensitive to variations in the overall IGMF strength. This signal is unavoidable in contrast to the inverse Compton pair halo/echo intensity, which is appealing in view of the large uncertainties on the IGMF in voids of large scale structure. It is also shown that the variability of blazar gamma-ray emission can be accommodated by the synchrotron emission of secondary products of UHE neutral beams if these are emitted by UHECR accelerators inside magnetised regions.}
	
\keywords{gamma-ray emission, ultra-high energy cosmic rays, extragalactic magnetic fields, propagation}
\maketitle

\section{Introduction}

Active Galactic Nuclei (AGN) are some of the very few types of astrophysical sources where ultra-high energy cosmic ray (UHECR) acceleration is expected to be viable. If proton acceleration does take place it may be possible to identify the signature of this process in blazar gamma-ray spectra depending on the total energy output in hadrons. Unambiguous detection of secondary gamma-rays of hadronic origin would have a lasting impact on the long standing question of the origin of UHECRs (see a review by e.g., \citealp{KO11}). 

BL Lacertae objects (BL Lacs) and flat spectrum radio quazars (FSRQs) commonly referred to as blazars are a radio loud subclass of AGN that have jets, i.e. collimated highly relativistic outflows, that point almost directly towards the observer \citep{Urry:1995mg}. Blazars appear point-like and most of them exhibit rapidly varying non-thermal emission across all energy bands thought to originate in the jet. FSRQs have broad prominent emission lines and are commonly associated to high-luminosity radio galaxies (Fanaroff-Riley type II [FR II]; \citealp{1974MNRAS.167P..31F}) whereas BL Lacs are associated with the lower luminosity Fanaroff- Riley type I [FR I] and have weak or absent emission lines. 

Blazar broadband spectral energy distributions (SEDs) have a double-bump shape. The low frequency peak is found in the optical to X-ray band and is generally thought to originate in the synchrotron emission of relativistic electrons and positrons in the jet. The high energy peak occurs in the gamma-ray band and its origin is less well understood. The most popular interpretation for the gamma-ray bump is synchrotron self compton (SSC) emission \citep{1997A&A...320...19M,1998ApJ...509..608T}. In this model it is assumed that relativistic electrons in the jet radiate in synchrotron producing photons up to X-ray energies which are subsequently IC scattered by the same electrons to gamma-ray energies. The model has the advantage of providing a simple explanation for the observed blazar spectrum based on a single emission region and of accommodating the rapid variability observed in the majority of blazars.

Along with electrons, protons may also be accelerated in blazars (see e.g., \citealp{Mannheim93}). High energy protons can lead to gamma-rays either via photo-hadronic pion production or Bethe-Heitler pair production, $p + \gamma \longrightarrow p + e^{+} + e^{-}$. In the case of heavier nuclei, photo-disintegration can also occur. Protons can also emit gamma-rays directly if they undergo synchrotron emission. Proton synchrotron requires strong magnetic fields of order $10-100$\,G and could occur in highly magnetised blobs inside the jet \citep{2000NewA....5..377A,2001APh....15..121M}. Irrespective of the mechanism responsible for the observed gamma-ray emission, the intrinsic blazar emission is absorbed and cascaded to lower energies due to interaction with the intervening Extragalactic Background Light (EBL) (e.g., \citealp{ACV94,Coppi:1996ze}).

In its five years of operation the Large Area Telescope ({\it Fermi}-LAT) on board the Fermi Gamma Ray Space Telescope \citep{Fermi-LAT:2011lqa} has detected and compiled a catalogue of more than 1000 blazars divided approximately equally into BL Lacs and FSRQs. Since 2004 when the current generation of ground based imaging atmospheric cherenkov telescopes (IACTs) began operations, more than 50 blazars have been detected in the TeV region\footnote{http://tevcat.uchicago.edu}. Most of the TeV blazars detected so far are classified as high-frequency peaked (HBL) within the blazar sequence \citep{Fossati:1998nw, Donato:2001ge}. A number of these sources exhibit TeV spectra that, after correcting for the expected absorption of the intrinsic emission by the EBL, are exceptionally hard (e.g., 1ES 1101-232, 1ES 0347-121, 1ES 0229+200, 1ES 1218+304, RGB J0710+591, see \citealp{HESS07_1101,HESS07_0347,Aharonian07,Albert06, Acciari09,Acciari10}). In order to explain their hard source spectra by electronic SSC or hadronic processes in inner jets, extreme source parameters and/or a special setup are required \citep{Aharonian:2005gh,Tavecchio11,Zacharopoulou11,Lefa11}. 

An alternative interpretation for these extreme blazar emissions is a proton-induced, intergalactic cascade emission. Ultra-high energy photons and pairs seeded by UHECRs via photo-meson and Bethe-Heitler pair production processes on the EBL are cascaded in the intergalactic medium \citep{Essey10,Essey13,Essey10b,2011APh....35..135E,Murase:2011cy,Tavecchio14}. If the magnetic field in the vicinity of the source is not strong then UHECRs will propagate away with little deflection in the intergalactic medium, where their products will initiate cascades much as they do in the case where the primary particles are leptons. However there will be differences in the observed gamma-ray spectra, since in the case of UHECR primaries the injection energies are much higher to start with and there is continuous injection of high energy electrons from Bethe-Heitler pair production, that leads to a tail of emission at high energies \citep{Essey10b,Murase:2011cy}. In particular, \citet{Takami:2013gfa} showed how this tail could be measured in the integrated flux above $\sim 500$\,GeV (depending on source redshift) for several luminous sources with $z \lesssim 1$ with the future CTA to disentangle between leptonic and the UHECR origin \citep[see also][for the capabilities of the approach for a statistical discrimination]{Zech:2013era}. Taking the same argument to extreme redshifts \citet{Aharonian13} showed that the discovery of TeV radiation from a source with $z \gtrsim 1$ could only be explained as this type of UHECR cascade emission with conventional physics (to date sources up to $z \lesssim 0.5$ have been observed at VHE).

Such cascaded emissions can produce slowly variable or almost non-variable components. In the presence of IGMFs the deflection that the cascade electrons will suffer in one cooling time, i.e. the mean timescale between subsequent energy loss events, can be estimated by considering the ratio of their Larmor radius, $r_{{\rm Lar}}$ at a given energy to the inverse Compton cooling distance, $D_{\rm IC}$, i.e. :  
\begin{equation}
\begin{centering}
\theta_e \sim \frac{D_{\rm IC}}{r_{{\rm Lar}}} = 1^{\circ} \left(\frac{B}{10^{-16}\,{\rm G}}\right) \left(\frac{E_e}{1~{\rm TeV}}\right)^{-2},
\label{eq:deflection}
\end{centering}
\end{equation}
where $D_{\rm IC} = 3m_ec^2/(4\sigma_{\rm T} U_{\rm CMB} \gamma_e) \simeq 690~{\rm kpc}(\gamma_e/10^6)^{-1}$ is the inverse Compton cooling length in the Thomson regime in terms of the CMB energy density $U_{\rm CMB}$. In the above expression $\sigma_{\rm T}$ is the electron Thomson cross-section and $\gamma_e$ the Lorentz factor of the electron. Therefore for the UHECR cascade channel to be observable weak IGMFs are required, with strengths in voids of order $B_{\rm voids}\, \lambda^{1/2}\lesssim 10^{-15}-10^{-14}\,{\rm G\,Mpc}^{1/2}$, with $\lambda$ the coherence length of the field. For stronger fields, the cascade radiation at TeV energies is suppressed by deflections of pairs in the IGMF. 

The detection of intergalactic cascade gamma-ray emission from gamma-ray bursts and AGN has also been proposed as a probe of IGMFs in the voids of large scale structure, for a pencil gamma-ray beam injection \citep[e.g.,][]{Plaga95,Dai02}. Given that if IGMFs are sufficiently strong the cascade radiation at TeV energies is suppressed by deflections of pairs one can derive lower limits on the IGMF from the non-observation of the otherwise expected inverse Compton cascade signal as demonstrated in e.g., \cite{Murase08b} quantitatively. Assuming that blazar emission is steady, {\it Fermi}-LAT data (or upper limits) lead to lower limits of $B_{\rm voids} \lambda^{1/2} \lesssim 10^{-15}~{\rm G}~{\rm  Mpc}^{1/2}$ (e.g., \citealp{Neronov10, Taylor11, Ahlers11, Vovk12}). Taking into account the possibility that blazar emission is transient rather than steady, conservative constraints can be obtained, which are order $10^{-18}-10^{-20}~{\rm G}~{\rm  Mpc}^{1/2}$ (e.g., \citealp{Murase08b,Dolag11, Dermer11}).

Upper limits from other studies constrain the mean IGMF strength to be below $B_{\rm mean}\, \lambda^{1/2}\lesssim10^{-8}\,{\rm G\,Mpc}^{1/2}$ \citep{Kronberg76, Blasi99}. However one expects the IGMF to be highly structured, with filaments supporting fields of $\sim 10^{-9}-10^{-7}$~G (e.g., \citealp{Ryu98}) and fields with strengths of the order of $10^{-7}-10^{-6}$~G in cluster regions (e.g., \citealp{Kim91,CKB01,CT02}). As discussed analytically in \cite{KAL11} and demonstrated by \cite{Murase:2011cy}, structured IGMFs play an important role by suppressing the resulting gamma-ray flux by more than one order of magnitude compared to the case without them. Note that this cascade component is not included in our calculations, as it has been studied extensively by the various authors cited above, and is not the aim of this paper. This channel is in any case expected to contribute mostly below $\sim 100\,$GeV, and we are primarily interested in studying the peak and tail of the observed VHE emission of a number of sources that challenge leptonic models. Besides, as discussed in Section~\ref{section:discussion}, the level at which this component contributes is very uncertain due to possible dilution in the IGMF.

On the other hand, if UHECR protons are accelerated in blazars that are embedded in magnetised or mildly magnetised regions they will inevitably give rise to electrons, typically carrying a few percent of the proton energy which will radiate synchrotron photons below the pair production energy, inhibiting the development of the inverse Compton cascade. The detectability of such signatures in a general framework has been studied analytically and by Monte-Carlo (concentrating on the synchrotron emission and without including the electromagnetic cascade component) by \cite{GA05}, \cite{Gabici:2006dv}, and \cite{2010JCAP...10..013K}. \cite{Prosekin:2011vf} studied analytically the synchrotron signature for UHECR sources located at high redshifts, when the energy conversion from protons to secondary particles was significantly more effective because of the denser and more energetic CMB in the past. \cite{Murase12} and \cite{Dermer12} investigated the highly variable synchrotron cascade emission, which can be caused by beamed UHE neutrals including photons and neutrons, taking into account the inverse Compton cascade component. \citet{AD08} studied the detectability of the gamma-ray spectra of UHECR seeded synchrotron emission, from the large scale jet of powerful AGN.

Here we demonstrate that this synchrotron pair halo/echo signal from UHECRs produced in blazars located in magnetised environments can also successfully reproduce the spectra observed in certain extreme hard-spectrum blazars. We show that this signal is unavoidable if blazars are UHECR sources embedded in structured regions, and that the energy flux at the peak energy is insensitive to the overall strength of the IGMF. We also point out that the variability of blazar gamma-ray emission can be accommodated by the synchrotron emission of secondary products of UHE neutral particles (neutrons and photons) if these are produced inside the source in accelerators of UHECRs.

\section{Methodology}

\subsection{Selection of sources}

The majority of VHE detected blazars are well described by the SSC model. Extreme TeV HBLs constitute a small, peculiar, fraction of the known blazar population which challenge such interpretation. We focus our study on a sample of the known extreme TeV HBLs with unusually hard VHE spectra for which there exist deep VHE observations. We are particularly interested in sources for which there exist long-term observations so as to constrain the variability properties. Namely we study the HBLs 1ES 0229+200, 1ES 1218+304 and RGB J0710+591 by way of demonstrating an alternative interpretation for the origin of the emission of this extreme HBL subpopulation. The sources studied here have been extensively discussed in other works (see references below) as they challenge leptonic models and a hadronic origin for their spectra has been discussed. The model presented in this work could apply to other extreme TeV HBLs, which have to be studied individually. 

The HBL 1ES 0229+200 \citep{Aharonian07} is a relatively distant, hard spectrum source. The observed spectral index $\Gamma_{\rm VHE} \simeq 2.5$ suggests a very hard intrinsic emission. Its high energy peak is at energy, $E \gtrsim 10$~TeV. Under the assumption that the intrinsic spectrum cannot be harder than $\Gamma \simeq 1.5$ in leptonic models, it has been inferred that the EBL must be very low, close to the minimum inferred by resolved galaxy counts in order to be consistent with the observed spectrum of 1ES 0229+200 \citep{Aharonian07}. The source has also been used to place lower limits on the strength of IGMFs in the voids of large scale structure, under the assumption of a purely leptonic origin of the gamma-ray emission. The gamma-ray spectrum of the source is also possible to explain if the emission is due to UHECR inverse Compton cascades \citep[e.g.,][]{Essey10, Murase:2011cy}. If the UHECR cascade mechanism is responsible for the observed spectrum, the lower limits of the IGMF inferred are no longer valid, nor are the EBL limits inferred from the study of this source. In addition, a number of non-conventional physics models have been considered in relation to its unusually hard spectrum \citep[e.g.,][]{Horns12}. There is a VERITAS campaign for deep observations of this object \citep{Dumm13}, where the first hints for variability in year-long timescales have been reported \citep{Aliu13}. If this variability is confirmed it will challenge the UHECR cascade origin of the observed VHE emission, as the secondary emission from this channel is not expected to exhibit variability at such timescales. There is no indication of variability of the spectral index of 1ES 0229+200 from observations so far. 
 
The HBL RGB J0710+591 was discovered in VHE by  \citep{Acciari10}. It has an observed index $\Gamma_{\rm VHE} \simeq 2.7$, making it one of the hardest VHE detected sources to date. There is no reported variability in the gamma-ray emission of this source. Its gamma-ray spectrum has been used to derive constraints on the IGMF strength assuming a leptonic origin \citep{Taylor11,Huan11,Arlen12}, whereas more recently a UHECR-induced inverse Compton cascade origin of the emission has been discussed \citep{Tavecchio14}. 

We also study 1ES 1218+304 \citep{Albert06, Fortin08}. Early observations of this source revealed no VHE variability, whereas a more recent deep observation campaign has revealed a $\sim 90$~hour variability in the VHE band \citep{2010ApJ...709L.163A}. As a result of it's unusually hard VHE spectrum, 1ES 1218+304 has also been used to set constraints on the IGMF strength \citep{Taylor11,Arlen12}. The main parameters of the three sources considered in this work are summarised in Table~\ref{tab:blazars}.

\begin{table}[h]
\centering
\begin{tabular}{lcrr}
&z & $\Gamma_{\rm VHE}$& $t_{\rm V}$\\
\hline
1ES 0229+200   & 0.14 & 2.5 & $\sim{\rm yr}$?\\ 
RGB J0710+591  &  0.13 & 2.7 &\mbox{none}\\
1ES 1218+304   & 0.18      &    3.0 &      $\sim 90\,{\rm hours}$\\ 
\hline
\end{tabular}
\caption{Redshifts $z$, observed spectral spectral index $\Gamma_{\rm VHE}$, and inferred variability $t_{\rm V}$ of the blazars studied in this work. (See text for references.)}\label{tab:blazars}
\end{table}

\subsection{Magnetic field model}

Besides limits on the mean strength of IGMFs and the values inferred in voids from observations as mentioned in the introduction, our knowledge on this subject is poor. A number of authors have modeled the large scale distribution of magnetic fields in the Universe with numerical simulations \citep[e.g.,][]{Ryu98, DGST05, Das08}. Unfortunately as shown in \citet{KL08b} these works do not converge as to the filling factor of magnetic fields especially at the weak magnetic field regime that one expects for IGMFs in voids. 

A more practical phenomenological method for modeling the structured magnetic fields in the Universe is to map the 
strength of magnetic fields $B$ to the underlying matter density distribution $\rho$ following analytic relations of the form $B = B_{0}f(\rho)$ where $B_{0}$ is a normalisation factor \citep{KL08a}, which roughly corresponds to the mean magnetic field strength in the universe. Here we consider a model: 
\begin{equation}\label{eq:B}
f_{\rm{contrast}}(\rho) = \rho \left[ 1 + \left(\frac{\rho}{\langle \rho \rangle}\right)^{-2} \right]^{-1}.
\end{equation}
The model is motivated by physical arguments related to the mechanism of amplification of magnetic fields during structure formation.  The relation (``contrast model'' of   \citealp{KL08a}) is an adhoc model which aims to capture a situation in which the magnetic field in the voids of large scale structure is suppressed with respect to the magnetisation in denser regions where magnetic pollution is expected to have taken place through eg. starburst, radio galaxies. We use this method to model the magnetic field in our simulation volume, setting $B_{0}$ of order 1\,nG. Note that the synchrotron emission calculated in this work depends very weakly on the chosen magnetic field model, as the major differences between the fields are in the voids, while the emission is produced in the dense magnetised regions. 

\begin{figure*}
 \centering
 \includegraphics[width=0.49\textwidth]{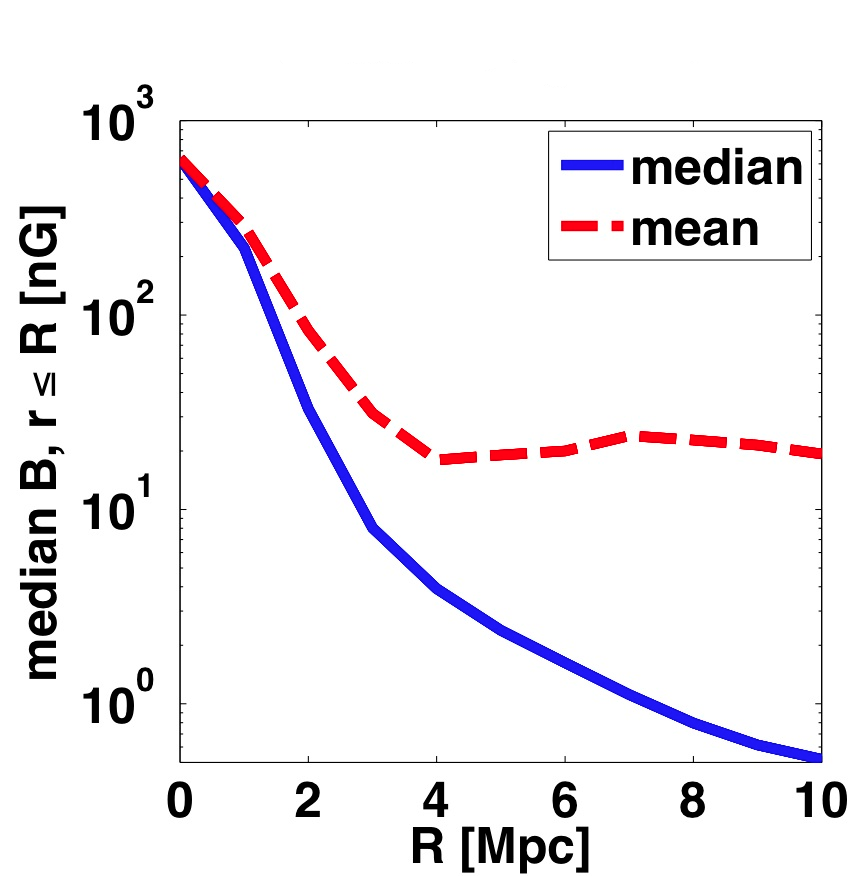}
 \includegraphics[width=0.49\textwidth]{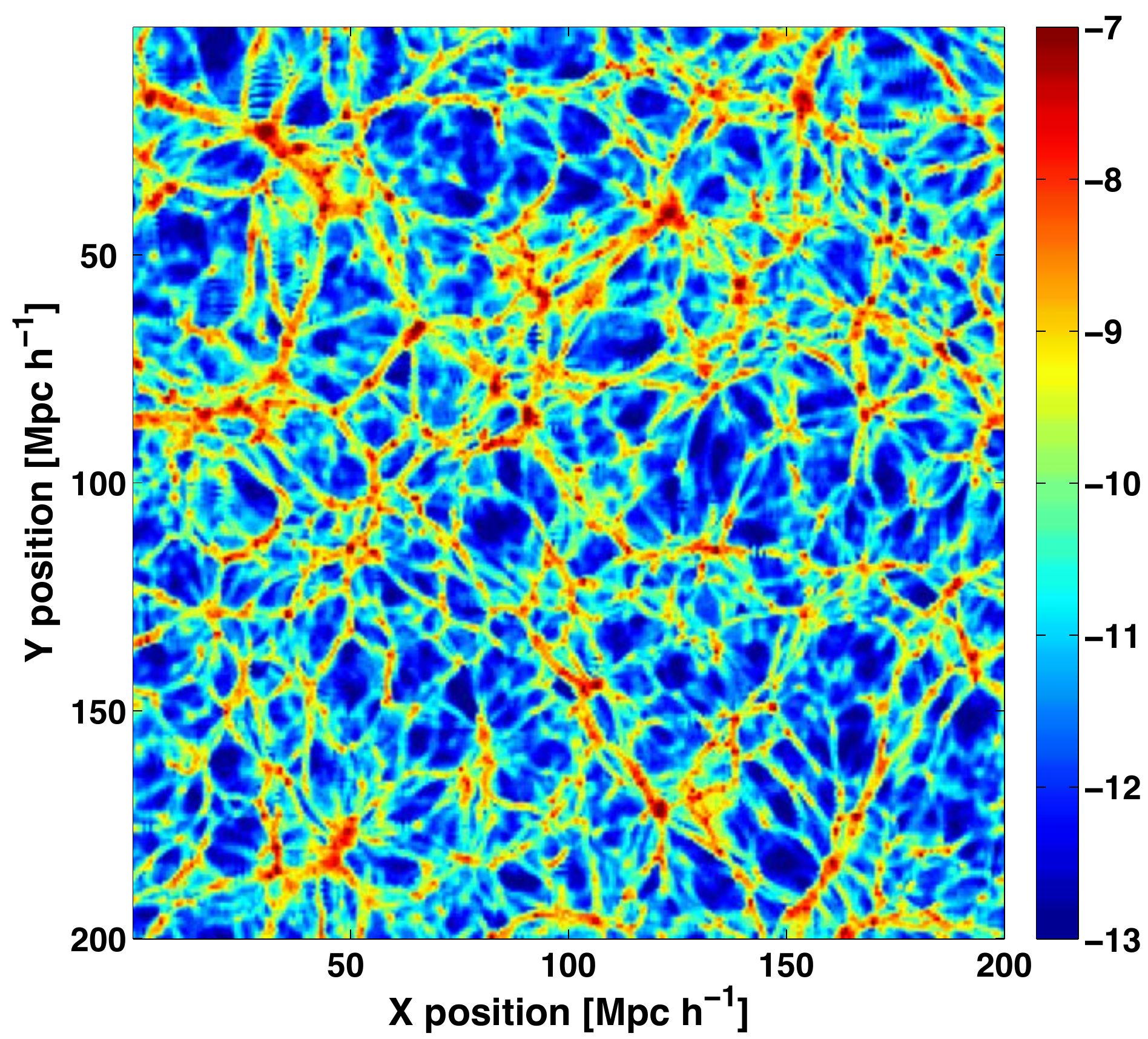}
  \caption{{\it Left panel:} Profile of the mean and median magnetic field strength as a function of distance from the source, in the vicinity of the source in our simulation volume. The profile shown corresponds to a normalisation factor $B_0 = 5$~nG (see text).
  {\it Right panel:} Slice through the plane of the source in our magnetic field simulation volume corresponding to $B_0 = 1$~nG.}  \label{fig:b-profile}
  \end{figure*}

To model the density field we use a dark matter simulation volume obtained with the hydrodynamical code RAMSES \citep{T02}. The distribution of dark matter is a good description of the underlying density field. The simulation volume is given by a $512^{3}$ cube which describes a volume of size $200\,h^{-1}$~Mpc. A standard $\Lambda \rm{CDM}$ model is assumed in the dark matter simulation and throughout this work with $\Omega_{\Lambda} = 0.7$, $\Omega_{\rm{M}} = 0.3$ and Hubble constant $H_{0} = 70~\rm{km}~\rm{s}^{-1}~\rm{Mpc}^{-1}$. Each cell of the simulation is considered as an independent scattering centre with coherence length $0.39\,h^{-1}$~Mpc with a randomly oriented magnetic field direction. A slice through the magnetic field volume in the contrast model can be seen on the right hand panel of Fig. \ref{fig:b-profile}.

\subsection{Simulations}
We model the expected gamma-ray signal through a combination of numerical Monte Carlo simulations and solving the kinetic equations of the particles in our simulations. For the propagation of hadrons emitted by the blazars in our simulation volume we use CRPropa 2.0 \citep{Kampert12} in 3D mode. CRPropa models the interactions of hadrons with the radio, CMB and IR backgrounds and subsequent production of electron/positron pairs, pions, neutrinos and electromagnetic cascades. The deflections and time delays of hadrons in cosmic magnetic fields are also handled. There are some limitations in the treatment of electromagnetic cascades with CRPropa and for this reason we use instead the code presented in \cite{Murase12} and \cite{Dermer12} solving the kinetic equations of the leptons in the simulations at every step, taking full account of energy losses and interactions with background photons. 

We inject protons in our simulations with a power-law spectrum of the form $dN/dE \propto E^{-\alpha}$ with luminosity chosen to reproduce the observed gamma-ray spectrum. In CRPropa pion production of protons on the CMB is simulated with the SOPHIA package \cite{Mucke99} and Bethe-Heitler pair production is treated following the parametrisation of \cite{Kelner08}. The equations of motion of the hadrons are solved at every time step, deflections are calculated analytically and a new position and momentum for each particle are determined. We record the positions and momenta of all the photons and electrons produced in photohadronic interactions in CRPropa. We collect the electrons and photons with a momentum vector that points towards a predefined observer at a specified distance from the source allowing a solid angle error margin of $2\pi (1-\cos{\theta})$, with $\theta = 11^{\circ}$ to increase statistics. We subsequently propagate the collected leptons with our kinetic code. 

In the lepton propagation code we inject the secondary leptons produced in the CRPropa runs uniformly over the magnetised region. This is a good approximation given that the loss length of protons to the relevant processes (Bethe-Heitler pair production and photo-pion production) is significantly larger than the size of the magnetised region. For the magnetic field strength, we take the volume averaged strength over the magnetised region out to 3\,Mpc (about the virial radius of an overdense region for clusters of galaxies; see Fig.~\ref{fig:b-profile} for an example of radial profile of the magnetic field) at this stage of the calculation. Quantitatively, with the structured field model presented in Eq.~(\ref{eq:B}), we set the normalization factor $B_0$ such that the volume averaged magnetic field strength over 3\,Mpc is $\bar{B} = 6,16,100,316$~nG respectively. In what follows, the term ``magnetised region" refers to the 3 Mpc radius sphere around the source, with mean field $\bar{B}$.
With this setup we compute the synchrotron signal that escapes from the magnetised region and apply a semianalytic cutoff for the attenuation of the signal that escapes the magnetised region by the intervening EBL. The synchrotron photons are below the pair production threshold for the typical magnetic field strength in filaments hence the gamma-ray photons observed in this case are prompt with no further deflections in the intergalactic medium. For the magnetic field strength typical in galaxy clusters some cascading of the TeV synchrotron photons might occur but this GeV-TeV inverse Compton component can be isotropised by IGMFs; as mentioned previously, the study of any such cascade contribution is not in the scope of this work. For the opening angle of the blazar jet we (arbitrarily) consider $\theta_{{\rm jet}} = 0.192 \simeq 11^{\circ}$, but the results presented here are not sensitive to this choice. 

For the EBL energy density and redshift evolution, we consider a range of models that are consistent with current observations \citep{Kneiske04, Kneiske08, Franceschini08, Inoue12} and for the CMB a black body spectrum of temperature 2.7~K. For the extragalactic radio background the model of \cite{Protheroe96} and measurements of \cite{Clark70} as implemented by CRPropa are used. Uncertainties on the spectrum and redshift evolution of the EBL and to a lesser extent of the radio background, introduce an uncertainty to our results but as we will show in the following section our results are robust to the choice of EBL model for the representative range of models that we have considered in this work. 

\section{Robustness of synchrotron signal with application to specific sources}
\label{sec:results}
The blazars studied in this work have gamma-ray peaks between $\sim$10 GeV$-$10 TeV, however irrespective of their intrinsic spectra a cut-off is observed in the TeV that strongly depends on the redshift of the source and details of the EBL spectrum and redshift evolution. The optical depth of the EBL to 1 TeV gamma-rays is thought to be $\mathcal{O}(1)$ at $z \sim 0.1$ hence for all the sources studied in this work a strong suppression of the intrinsic source flux above this energy is expected.

\begin{figure}
\centering
\includegraphics[width=0.49\textwidth]{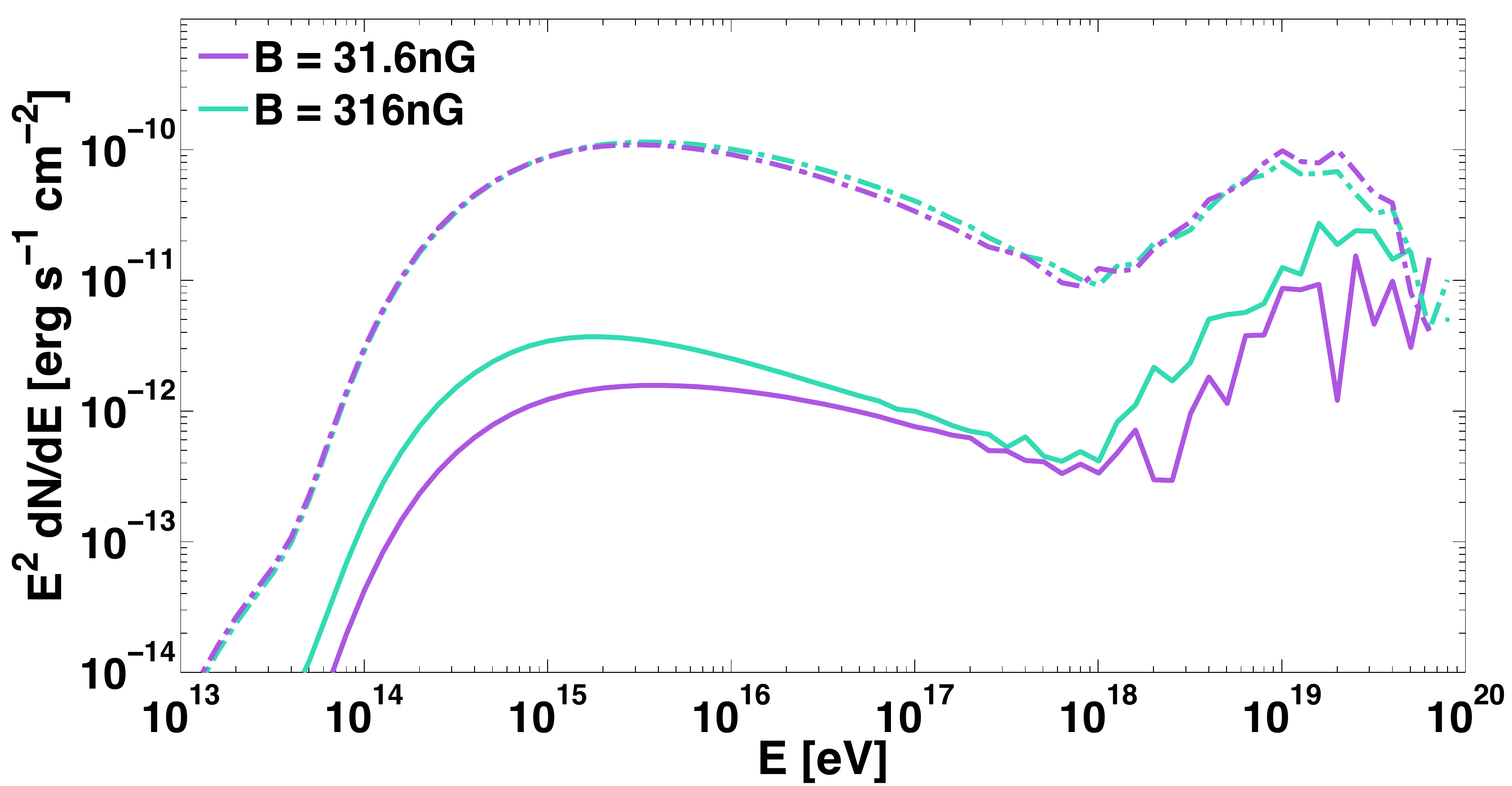}
\caption[The energy flux of secondary leptons produced by $p \gamma$ interactions in a 3 Mpc magnetised region around a UHECR source at redshift $z = 0.14$.]{Solid lines: The energy flux of secondary leptons produced by $p \gamma$ interactions in a 3 Mpc magnetised region around a source at redshift $z = 0.14$ that emits UHECRs with $L_{{\rm cr,iso}} = 10^{47}{\rm erg}~{\rm s}^{-1}$. The volume averaged magnetic field strength in the magnetised region is $\bar{B} = 31.6~{\rm nG}~({\rm purple}), 316~{\rm nG}~({\rm green})$. The noise in the pion bump is due to the finite number of particles injected in our simulations. Dot-dashed lines: The energy flux of secondary leptons produced by $p \gamma$ interactions beyond the first 3 Mpc of propagation.}
\label{fig:secondary_flux_level}
\end{figure}

In the secondary synchrotron model the main contribution to the secondary energy flux within the magnetised region will be from photomeson production due to the significantly shorter cooling length compared to that of Bethe-Heitler pair production. Fig.~\ref{fig:secondary_flux_level} presents the secondary leptons (photons and electron-positron pairs) produced inside and outside the magnetised region for a source at redshift $z = 0.14$ that emits UHECRs with $L_{{\rm cr,iso}} =\int_{10^{18} \rm{eV}} {\rm d}E ({\rm d} L_{\rm cr,iso}/{\rm d}E) = 10^{47}{\rm erg}~{\rm s}^{-1}$. Here and throughout the injected luminosity quoted is above $10^{18}$~eV and the injection spectral index, $\alpha = 2.0$, unless otherwise stated. Protons with energy lower than $\sim 10^{18}$~eV should be present in the jet and will contribute to the total jet power but not the observed gamma-ray flux as they are most likely confined in the jet. 
Considering the contribution of protons with $E_{\rm min} \gtrsim \Gamma m_p c^{2}$ to the total jet power, where $\Gamma \simeq 10$ is the typical Lorentz factor of the bulk motion and $m_p$ the proton mass, increases the $L_{{\rm cr,iso}}$ required to produce the same secondary lepton flux by $\ln(E_{\rm max}/E_{\rm min})/\ln(E_{\rm max}/{10}^{18}~{\rm eV})$, which is a factor of a few. We observe in Fig.~\ref{fig:secondary_flux_level} that for our chosen injection spectrum the peak of the energy spectrum of the first generation of electrons from photomeson production is at $E_{e} \sim 10^{19}$~eV as a result of the competition between the abundance of primary protons with increasing energy and the energy loss rate of the primary protons. The characteristic energy of the synchrotron emission of these electrons will be at $E_{\rm syn} \sim 6.7 \times 10^{11} (B/100~\rm{nG})$~eV, which for the typical magnetic fields expected in the large scale structures we study, is near the peak of the blazar spectra. The synchrotron emission which is emitted with energy beyond few TeV will be absorbed by the EBL. The dot-dashed component in Fig.~\ref{fig:secondary_flux_level}, produced beyond the first 3 Mpc from the source, could also contribute to the cascaded emission, as its level is higher than the flux produced closer to the source. As already mentioned this low energy component is likely to be diluted by IGMFs and not contribute to the GeV flux of the source if intervening IGMFs are non-negligible. In this sense the results shown here correspond to the limit where IGMFs are strong enough to isotropise this low energy cascade component. 
\subsection{1ES 0229+200}

\begin{figure}
\centering
\includegraphics[width=0.49\textwidth]{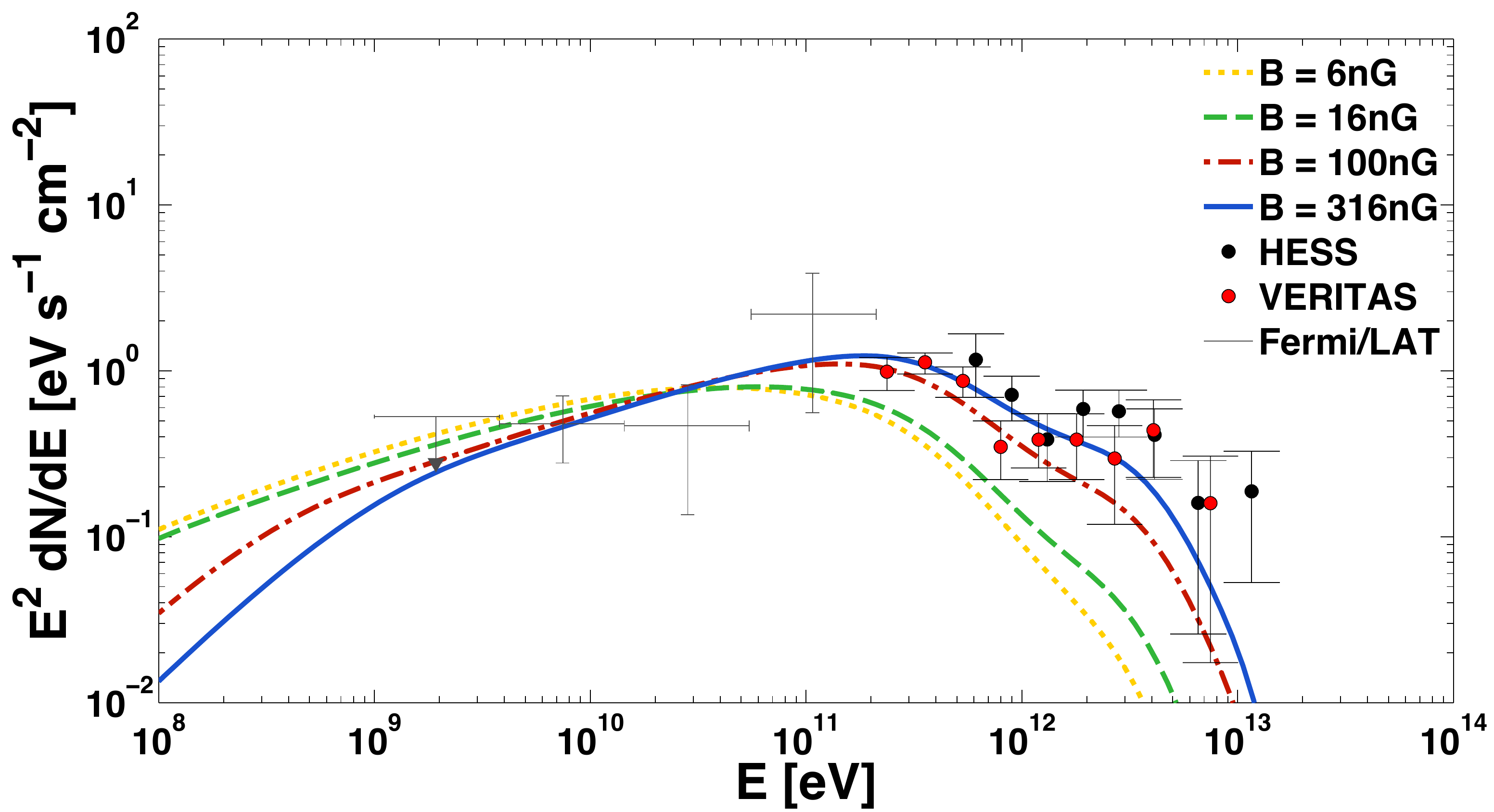}
\caption[The fit of the UHECR secondary synchrotron model to the spectrum of 1ES 0229+200 for a range of values of the volume averaged magnetic field strength in the magnetised region surrounding the accelerator.]{The fit of the UHECR secondary synchrotron model to the spectrum of 1ES 0229+200, assuming a mean strength of the magnetic field in the magnetised region, $\bar{B}$ in the range $6-316$~nG. Here $L_{{\rm cr,iso}} = 10^{46.5}{\rm erg}~{\rm s}^{-1}$ has been assumed. {\it Fermi}-LAT data points for this source here and throughout have been adapted from \citet{Vovk12}. The model spectra shown account for the attenuation by the EBL, for which the model of \citet{Kneiske08} has been considered.}
\label{fig:var_egmf}
\end{figure}

Fig.~\ref{fig:var_egmf} shows the model prediction of the secondary synchrotron signal to the observed spectrum of 1ES 0229+200 for $\bar{B}$ in the range $6-316$~nG. The assumed isotropic equivalent luminosity is $L_{{\rm cr,iso}} = 10^{47}{\rm erg}~{\rm s}^{-1}$. For this source, whose spectrum peaks at $\gtrsim 10$~TeV, $\bar{B} = 316$~nG is consistent with the combined GeV-TeV data, whereas considering values of $\bar{B}\lesssim 100\,$nG results in a poorer fit. 

\begin{figure}
 \centering
 \includegraphics[width=0.49\textwidth]{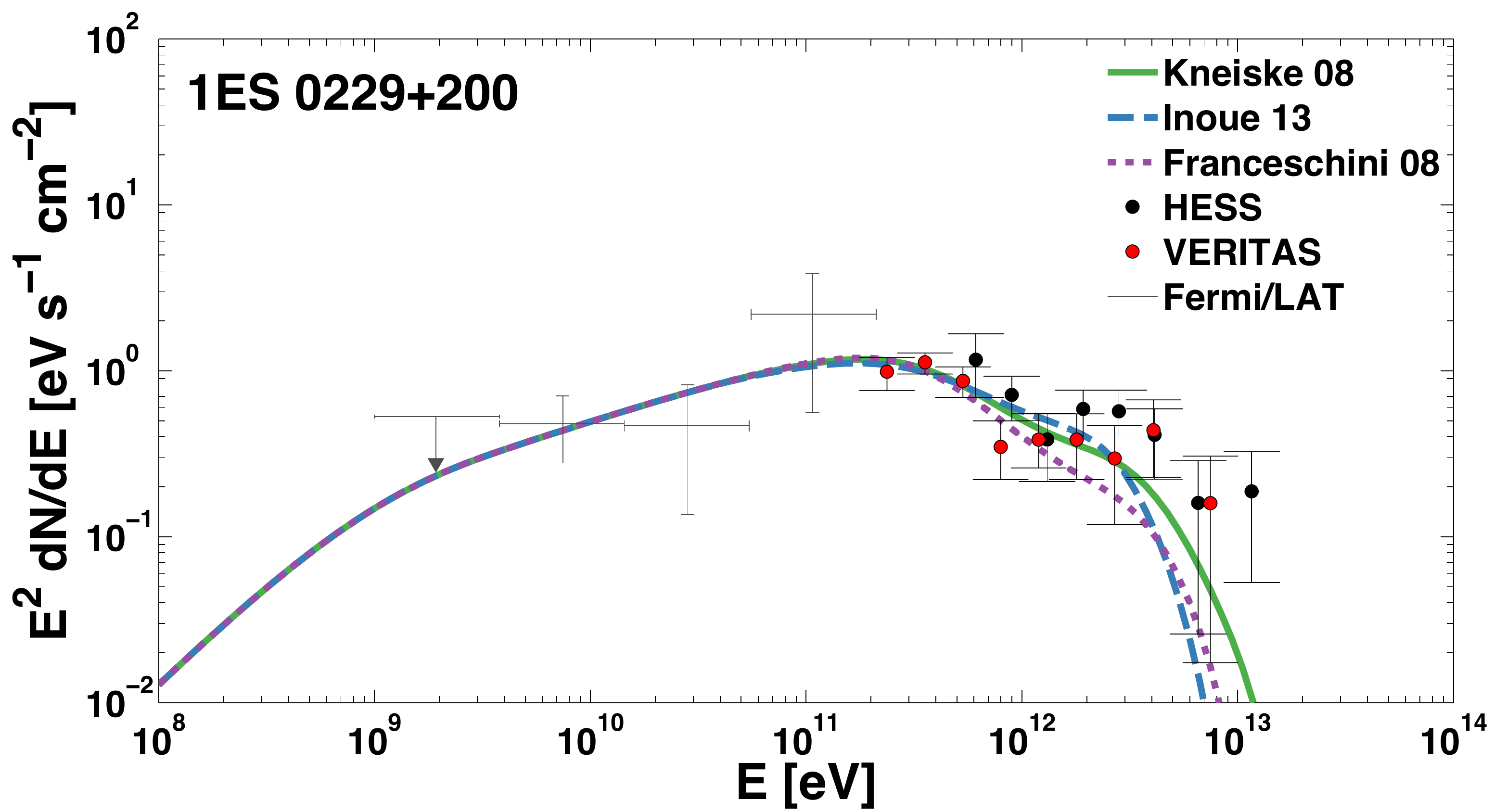}\\
  \caption[Robustness of the UHECR secondary electron synchrotron model fit to the spectrum of 1ES 0229+200 for a selection of EBL models.]{Robustness of the UHECR secondary electron synchrotron model to the EBL model considered for 1ES 0229+200. Here we show results with the EBL model of \citet{Franceschini08}(purple dotted), \citet{Kneiske08} (green solid), \citet{Inoue12} (blue long dashed). Model parameters assumed are $\bar{B} = 316$~nG and $L_{{\rm cr,iso}} = 10^{46.5}{\rm erg}~{\rm s}^{-1}$.}
   \label{fig:Robustness_to_EBL}
  \end{figure}

In Fig.~\ref{fig:Robustness_to_EBL} we show the robustness of the model fit to the uncertainty in the intensity and redshift evolution of the EBL, by considering a range of EBL models that are consistent with existing limits and measurements. The goodness of the model fit to the spectrum of 1ES 0229+200 depends on the EBL assumed and the best fit is obtained with the lower limit model of \citet{Kneiske08}. All models considered slightly under-predict the energy flux at the highest TeV datapoint but for the fit with the EBL model of \citet{Kneiske08} this disagreement is very small. Considering a slightly higher value of $\bar{B}$ would improve the consistency of the model with the last TeV data-point. 

\begin{figure}
 \centering
 \includegraphics[width=0.49\textwidth]{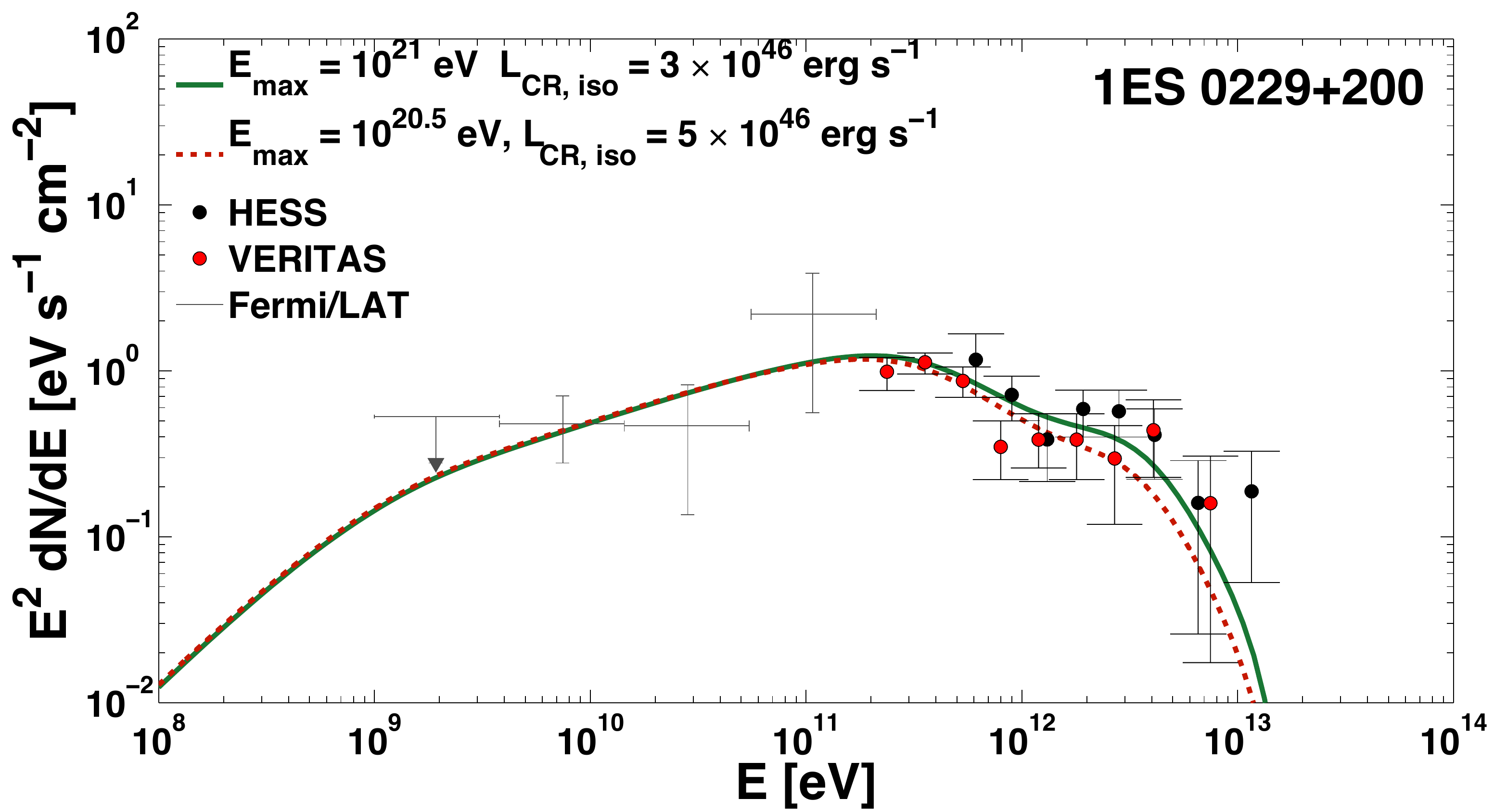}\\
 \includegraphics[width=0.49\textwidth]{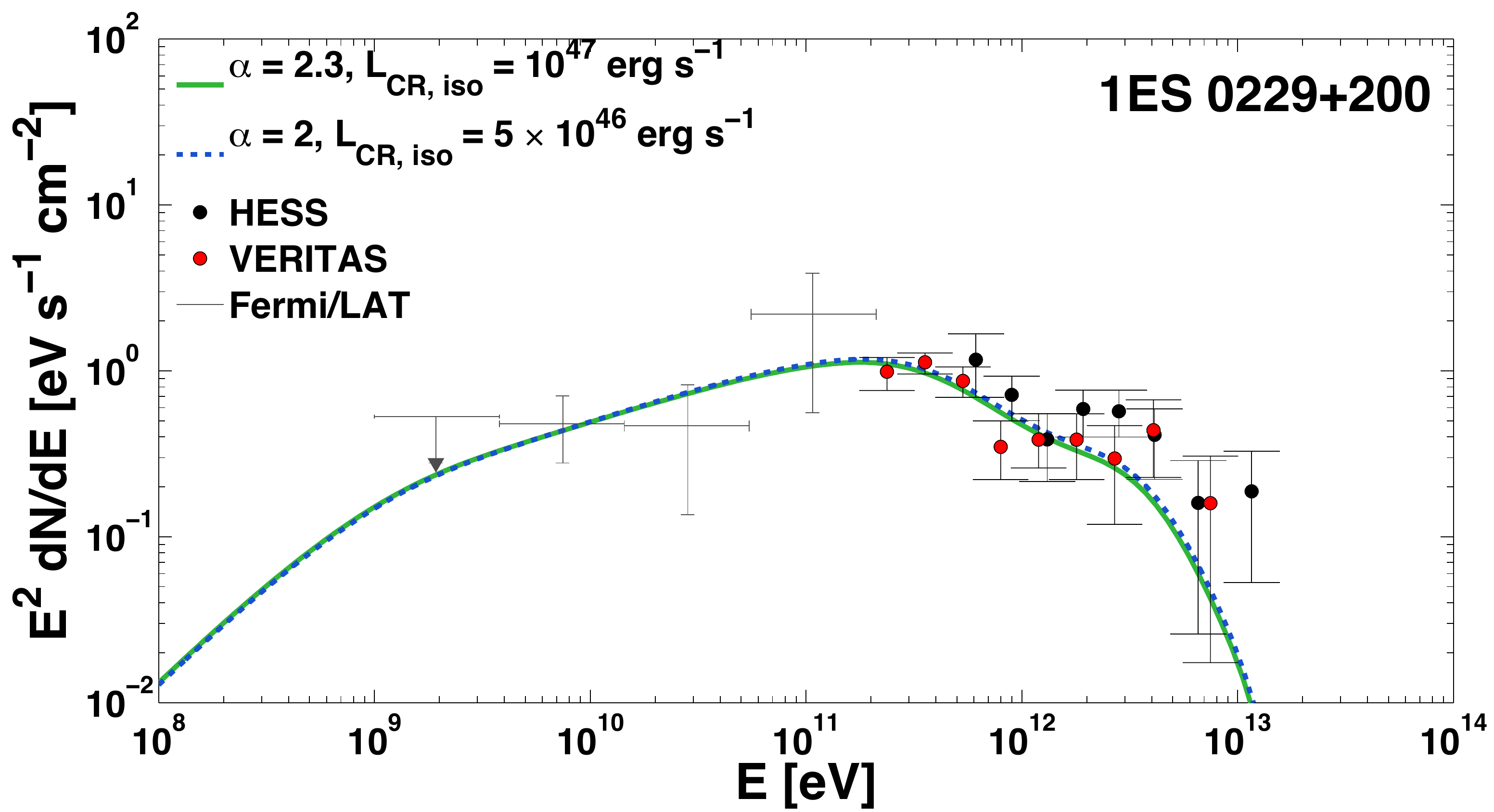}\\
  \caption[Robustness of the UHECR secondary electron synchrotron model fit to the spectrum of 1ES 0229+200 for different UHECR injection parameters.]{Robustness of the UHECR secondary electron synchrotron model fit to the spectrum of 1ES 0229+200, for different UHECR injection spectrum parameters. Here $\bar{B} = 316$~nG has been assumed.\\
 {\it Top panel:} Model prediction for $E_{{\rm max}} = 10^{20.5}~{\rm eV}$ (red dotted line) and $E_{{\rm max}} = 10^{21}~{\rm eV}$ (green solid line). {\it Bottom panel:} Model prediction for injection spectral index $\alpha = 2$  (blue dotted line) and $\alpha = 2.3$ (green solid line).}
\label{fig:Robustness_to_emax}
\end{figure}

In Fig.~\ref{fig:Robustness_to_emax} we show the robustness of the model predictions to the choice of UHECR injection parameters, namely the maximum UHECR injected energy $E_{{\rm max}}$ and spectral index $\alpha$. In the top panel we show the model fit to the spectrum of 1ES 0229+200 for $E_{{\rm max}} = 10^{20.5}$~eV, $10^{21}$~eV. Here for $\bar{B}$ we have assumed $316$~nG and $L_{{\rm cr,iso}} \sim 10^{46}-10^{47}~{\rm erg}~{\rm s^{-1}}$ (see exact value in the figure legend for each case studied). Lower values of $E_{\rm max}$ result in a poorer fit, implying that if the synchrotron channel presented in this work is at the origin of the measured blazar spectrum, then these accelerators can reach the highest observed energies of cosmic rays. Considering a higher value of $E_{{\rm max}}$ results in a harder model spectrum, increasing the consistency to the highest energy TeV data-point while easing the luminosity requirements on the accelerator. This is as expected intuitively as higher energy UHECRs undergo $p\gamma$ interactions quicker thus creating more flux in the first few megaparsecs of propagation. In the bottom panel, the model prediction for the spectrum of 1ES 0229+200 for $\alpha = 2.0, 2.3$ is shown. A softer spectrum increases the required $L_{{\rm cr,iso}}$ but the model prediction remains otherwise unaltered. Note that the required $L_{{\rm cr,iso}}$ is very similar to the required UHECR luminosity in the UHECR-induced inverse Compton cascade scenario for this source if it is embedded in a filament \citep{Murase:2011cy}.

\subsection{Other sources}

\begin{figure*}
\centering
\includegraphics[width=0.49\textwidth]{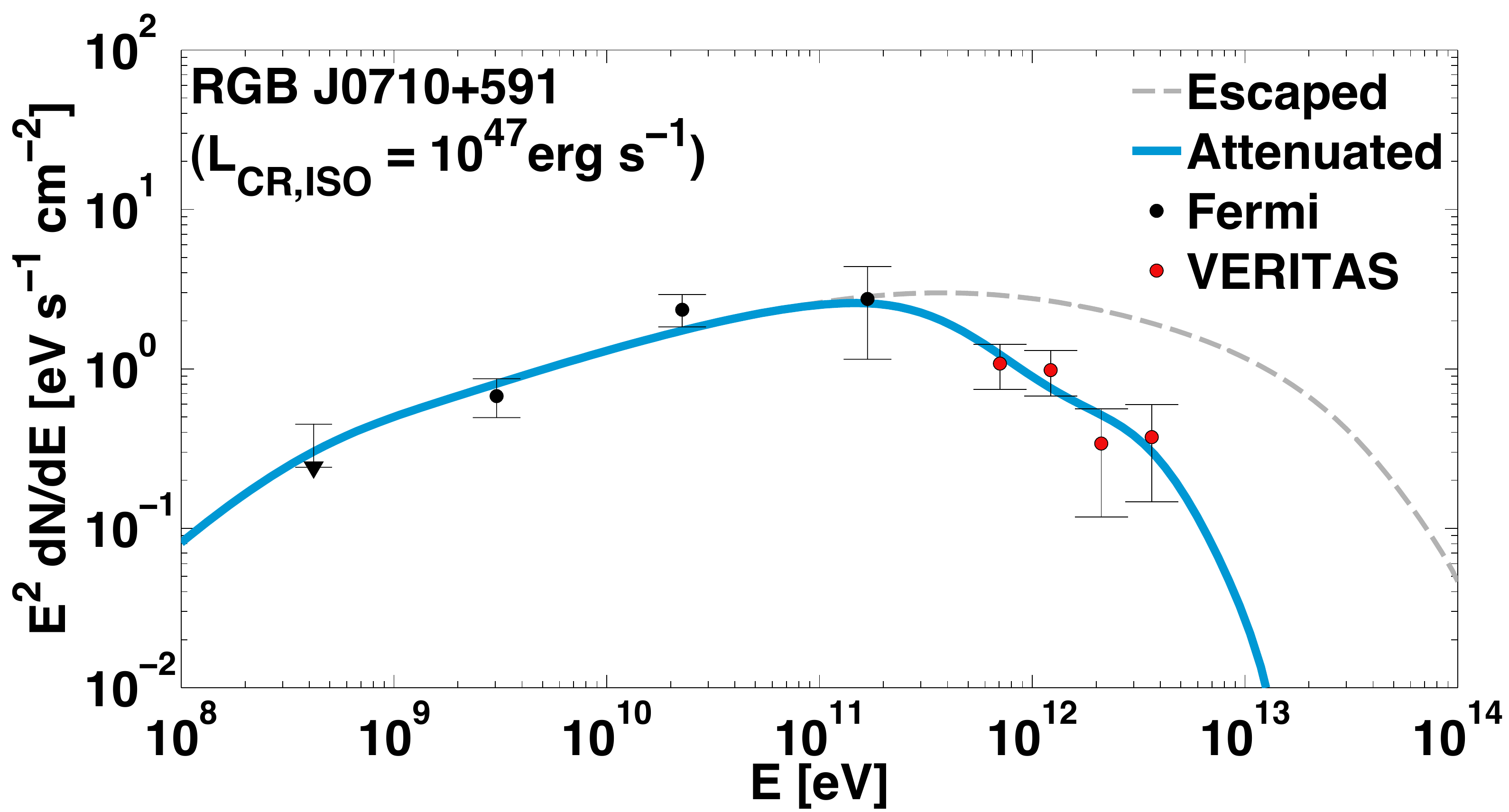}
\includegraphics[width=0.49\textwidth]{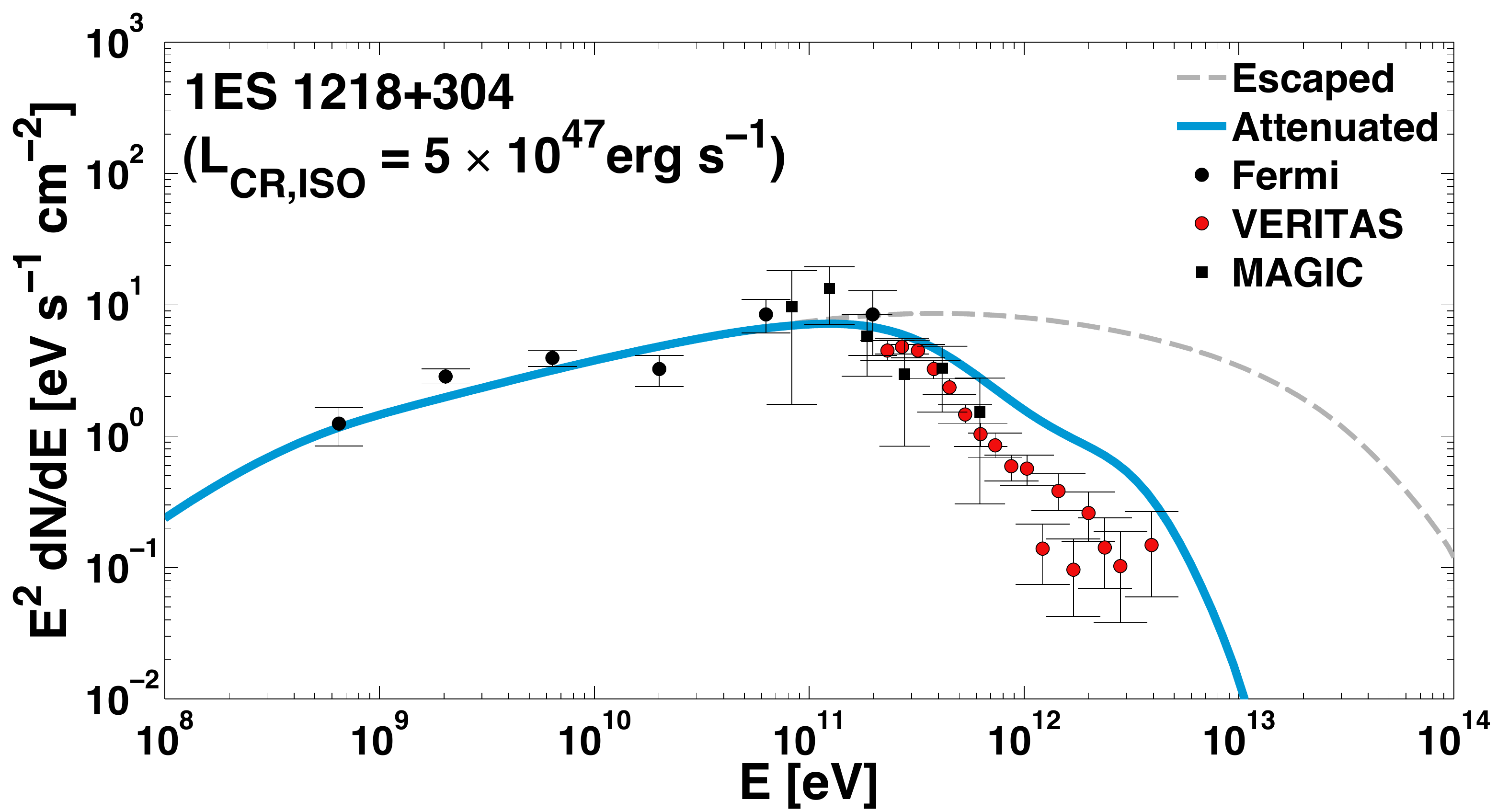}
\caption{The expected photon energy flux, resulting from the UHECR secondary electron synchrotron model a magnetised region with average magnetic field strength $\bar{B} = 100$~nG for RGB J0710+591 (left) and 1ES 1218+304 (right). Long-dashed lines show the spectrum that escapes from the magnetised region. Blue solid lines show the spectra with account of the attenuation by the EBL, for which the model of \citet{Kneiske08} has been considered.}
\label{fig:other_sources}
\end{figure*}

The fits to the combined GeV-TeV band spectra found in the UHECR secondary synchrotron model under the assumption of  $\bar{B} = 100$~nG in a 3 Mpc magnetised region around the source are shown in Fig.~\ref{fig:other_sources} for RGB J0710+591 and 1ES 1218+304. For RGB J0710+591 the synchrotron pair echo model is consistent with the data across the GeV and TeV bands and $L_{{\rm cr,iso}}=10^{47}{\rm erg}~{\rm s}^{-1}$. 

The fit to the TeV data for 1ES 1218+304 is poorer although the model is consistent with the GeV observations for this source. Considering a slightly lower value of $\bar{B}$ and/or a softer injection spectrum would fit the TeV data, at the cost of a slight increase in the required $L_{{\rm cr,iso}}$ as demonstrated in Fig.~\ref{fig:var_egmf} for 1ES 0229+200. However as already noted the gamma-ray emission from this source exhibits a variability in $\sim$~day timescales. The observed variability does not favour a secondary UHECR synchrotron origin of the spectrum of this source (see discussion in section \ref{subsec:uhephotons}). 
\subsection{UHE neutrals}
\label{subsec:uhephotons}

If UHECRs are accelerated in AGN, $p\gamma$ interactions \textit{inside} the source should also lead to the production of UHE photons and UHE neutrons \citep{Murase09b}. Such UHE neutral particles can leave the acceleration zone if the target photon spectrum is thermal or suppressed by synchrotron self-absorption. In particular, if the radio emission from bubbles and lobes is not too strong, the mean free path to $\gamma \gamma$ pair production is $\sim 2$~Mpc, which is comparable to the size of the magnetised region considered here. So about a half of the UHE photons can escape while the other half would promptly produce electron-positron pairs in the magnetised intergalactic medium surrounding a source such as we have been discussing in this work. The secondary high energy electrons would in turn radiate synchrotron photons as in the case of runaway UHECRs we have considered so far. This type of emission has recently been proposed and studied by \citet{Murase12}.

One expects the signature of the secondary synchrotron emission of runaway UHE photons to differ from the emission we have studied thus far of runaway UHECR protons in a number of ways. Firstly the injection spectra in the two channels are different; in the case of runaway UHECRs the only relevant photon fields are the cosmic photon backgrounds whereas the production spectrum of UHE photons must happen on a target photon field inside the source. \citet{Murase12} analytically calculated the expected UHE photon spectrum considering target photon spectra created by the synchrotron emission of electrons in the source environment. Further, and more importantly the timing properties of the signal should be different between the two channels. We explore this further in the following section.

\begin{figure}
\centering
\includegraphics[width=0.49\textwidth]{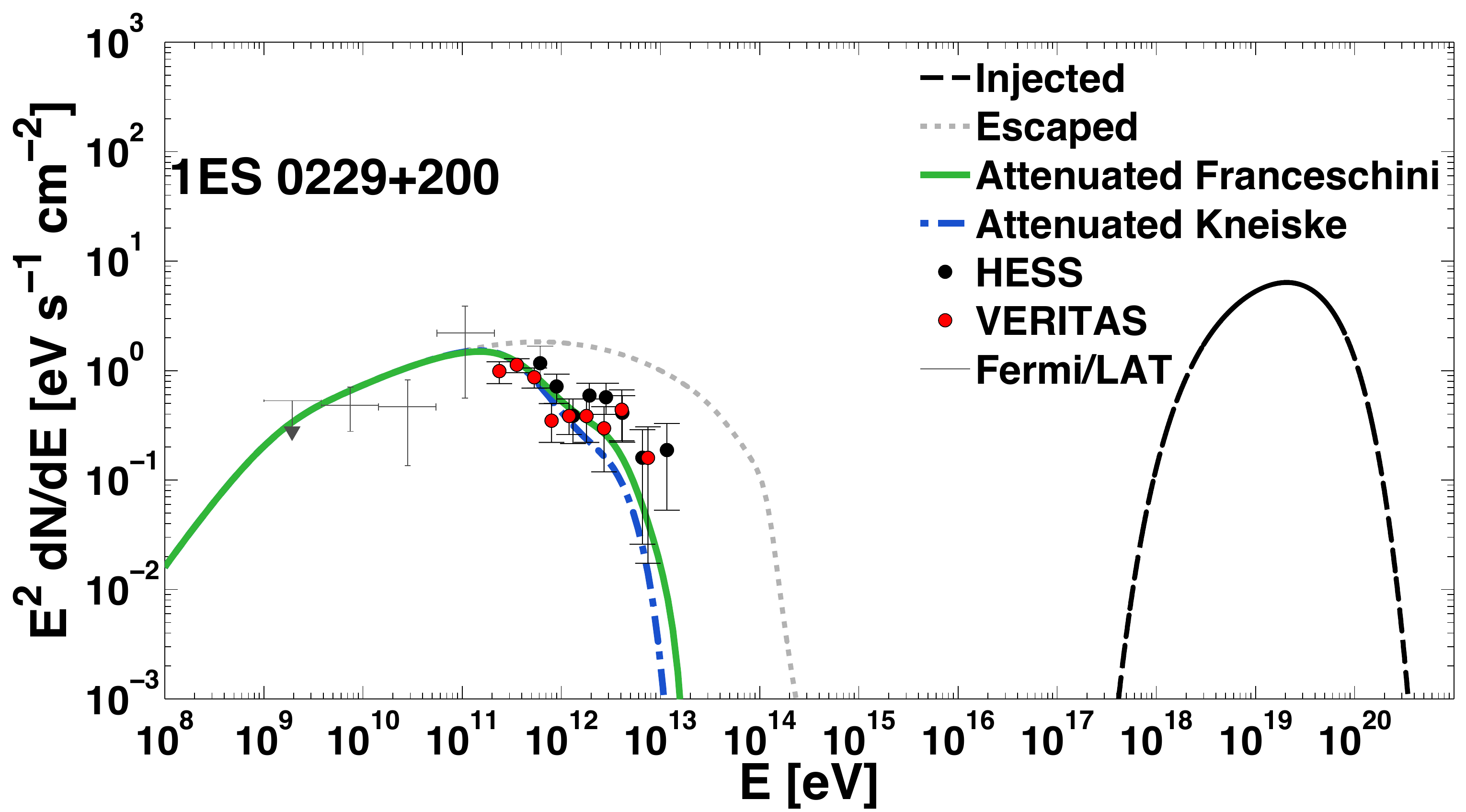}
\caption{The arriving energy flux expected from the UHE photon emission escaping from a magnetised region with average magnetic field strength $\bar{B} = 316$~nG for 1ES 0229+200. The injected luminosity normalisation in UHE photons is $L_{0} = 10^{45}{\rm erg}~{\rm s}^{-1}$ (see text for details). The dashed black line gives the injected UHE photon spectrum, the grey dotted line shows the spectrum that escapes from the magnetised region. The green solid line and blue dot-dashed line show the expected attenuated spectra using the EBL model of \citet{Kneiske08} and \citet{Franceschini08} respectively.} 
\label{fig:UHE_photons_with_injection}
\end{figure}

\begin{figure}
\centering
\includegraphics[width=0.49\textwidth]{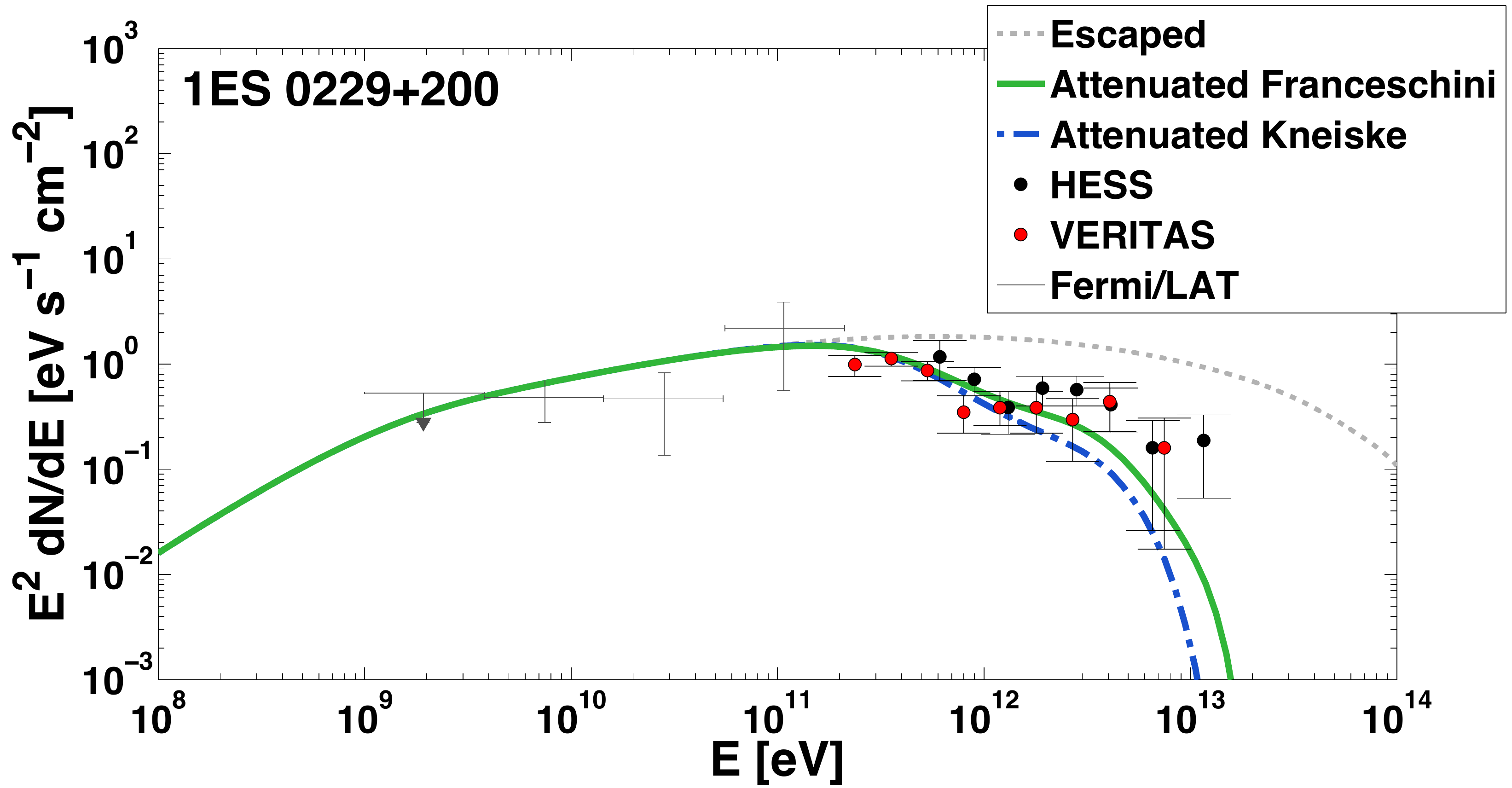} 
\includegraphics[width=0.49\textwidth]{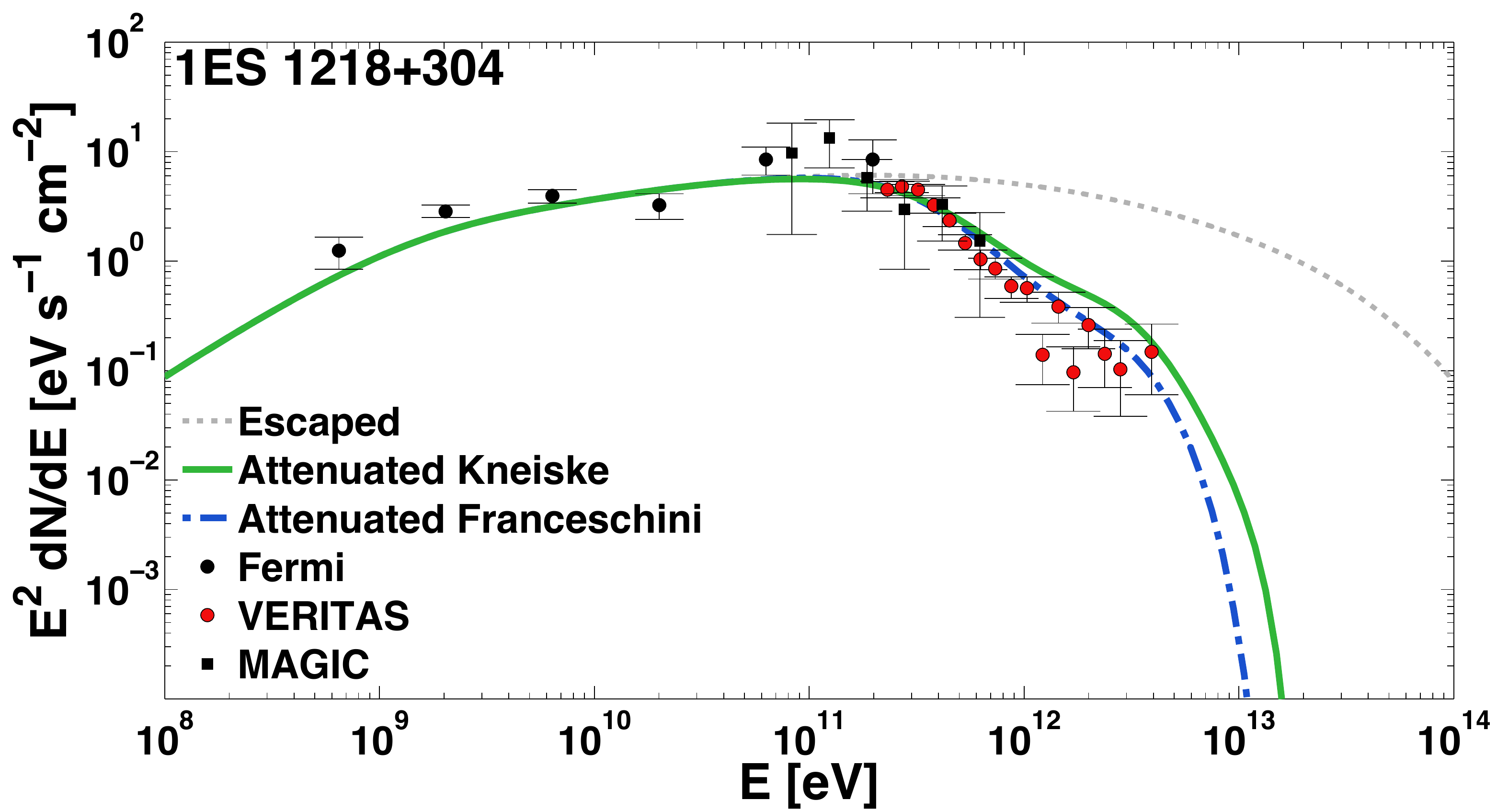}
\caption{{\it Top panel:} Same as Fig.~\ref{fig:UHE_photons_with_injection} but zooming in at the arriving photon energy flux. The volume averaged magnetic field strength inside the magnetised region is assumed to be $\bar{B} = 316$~nG. {\it Bottom panel:} Same as on the top panel but for 1ES 1218+304. The injected luminosity normalisation is $L_{0} = 8 \times 10^{45}{\rm erg}~{\rm s}^{-1}$ and the volume averaged magnetic field strength inside the magnetised region is assumed to be $\bar{B} = 100$~nG.} 
\label{fig:UHE_photons}
\end{figure}

In figures \ref{fig:UHE_photons_with_injection} and \ref{fig:UHE_photons} we show the expected gamma-ray spectra from the UHE photon channel for 1ES 0229+200 and 1ES 1218+304 motivated by their observed variability (or hints of such a variability in the case of 1ES 0229+200). Following \citet{Murase12} we have considered an injection spectrum of the form $L_{{\rm \gamma}} = L_{0} \times (E/E_{\gamma}^{{\rm max}})^{0.5}e^{-E_{\gamma}^{{\rm min}}/E}e^{E/E_{\gamma}^{{\rm max}}}$. Here the generation spectrum of UHE photons depends on the slope of the primary proton spectrum $\alpha$, as well as the slope of the target photon spectrum $\zeta$ as $E_{\gamma}^{2} {\rm d}\gamma/{\rm d}E_{\gamma} \propto E_{p}^{1+\zeta-\alpha}$. We have taken $\zeta \sim 1.5$ which is typically expected for a photon field generated via synchrotron emission in AGN and $\alpha \simeq 2.0$ as throughout most of this work. The values of $E_{\gamma}^{{\rm min}} = 10^{18.5}$~eV and $E_{\gamma}^{{\rm max}} = 10^{19.5}$~eV are chosen to capture the typical energies of the UHE photons that are created through the $p\gamma$ interaction, corresponding to maximum proton energy $E_p^{\rm max} = 10^{20.5}$~eV. The normalisation $L_{0} \sim f_{p \gamma} L_{{\rm cr, iso}}$, where $ f_{p \gamma}$ is the efficiency with which UHE photons are produced in $p \gamma$ interactions. For the setup considered here we take $f_{p \gamma} \sim 1/200$ at $E_p^{\rm max}$. For the UHE photon generation spectrum, see \citet{Murase12} and references therein. Here we show some examples of this channel for demonstration purposes. 

The model prediction is consistent with the GeV-TeV data of 1ES 0229+200.  As discussed above, a slightly higher value of the magnetic field strength at the source would shift the peak of the synchrotron emission further into the TeV providing consistency with the TeV observations for this source. The required luminosity is $L_{0} = 10^{45}{\rm erg}~s^{-1}$ i.e. $L_{{\rm cr, iso}} = 2 \times 10^{47}{\rm erg}~s^{-1}$, similar to the required UHECR luminosity in the UHECR-induced synchrotron cascade.

Comparison of figures~\ref{fig:Robustness_to_emax}, \ref{fig:other_sources} and \ref{fig:UHE_photons} illustrates the differences between the two channels we have studied. A harder spectrum is observed in the UHECR channel for a given magnetic field strength. The injection of a spectrum with an exponential cut-off in the UHE photon channel partly explains why the resulting synchrotron spectrum in this model has a sharper cut-off than the synchrotron spectrum in the UHECR channel. A further difference comes from the contribution of the Bethe-Heitler process in the UHECR channel. The secondary pairs that are created via Bethe-Heitler pair production contribute to the spectrum that escapes the magnetised region through the addition of photons with energy beyond $\sim 10^{14}$~eV via inverse Compton scattering. This is because the Bethe-Heitler component, which peaks at $\sim 10^{15}$~eV is below the critical energy for cooling via synchrotron emission which is otherwise the dominant cooling mechanism in the magnetised regions we have been considering. This contribution from Bethe-Heitler pairs results in a harder spectrum escaping the magnetised region in the UHECR channel. In this sense the observation of the UHE photon case resulting in a softer TeV spectrum should not be considered a general result, the difference comes from the different initial conditions. 

\begin{figure}
\centering
\includegraphics[width=0.49\textwidth]{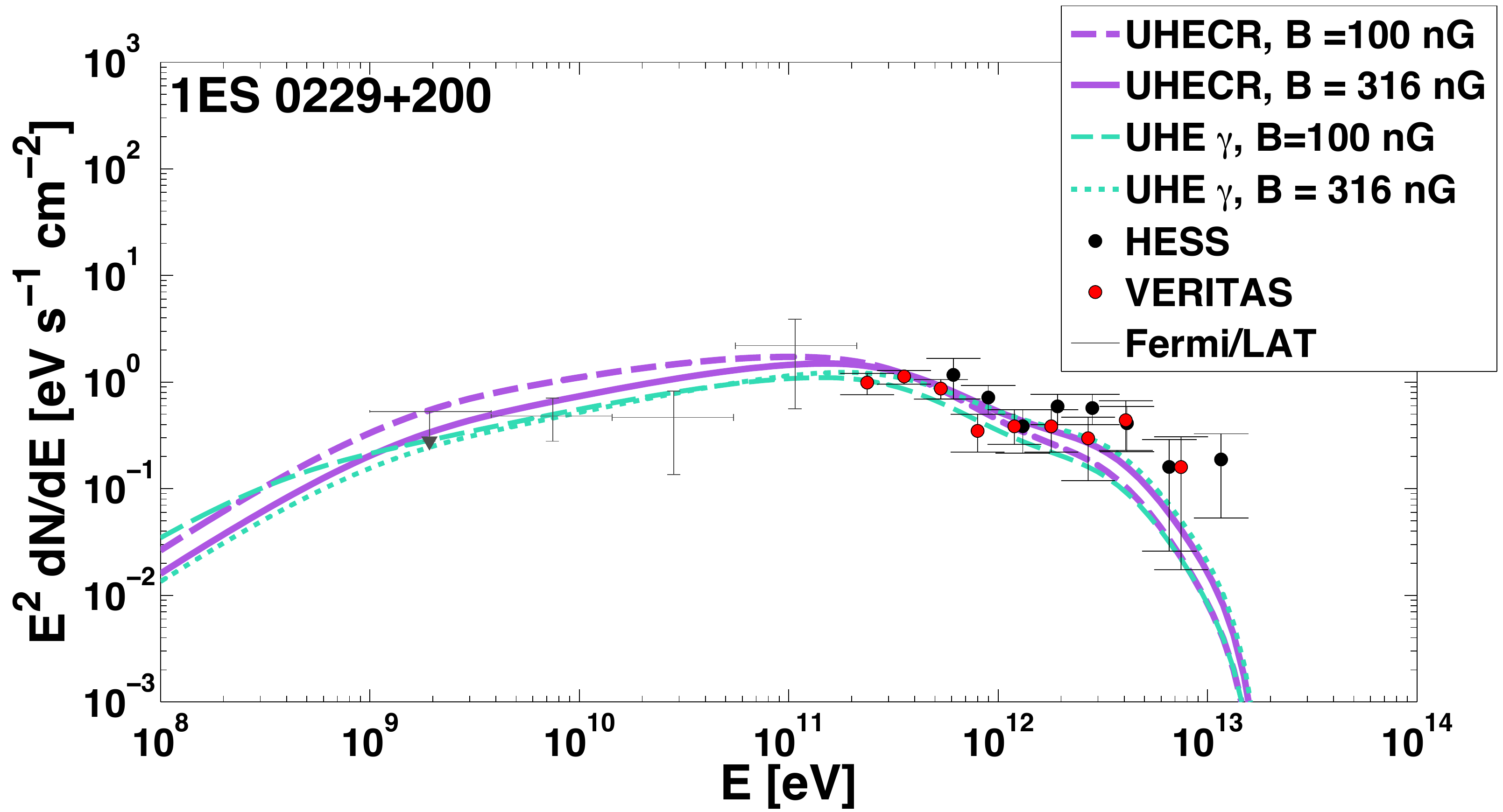} 
\caption{Comparison of the expected energy flux of secondary leptons from UHECR primaries (purple lines) and UHE photons (light blue lines) for a magnetised region with $\bar{B} = 100, 316$~nG for 1ES 0229+200. In the UHECR channel the injected luminosity is $L_{{\rm cr,iso}} = 10^{46.5}{\rm erg}~{\rm s}^{-1}$. In the UHE photon channel the injected luminosity normalisation is $L_{0} = 8 \times 10^{45}{\rm erg}~{\rm s}^{-1}$.}
\label{fig:photons_protons}
\end{figure}

Fig. \ref{fig:photons_protons} compares the signatures of runaway UHECRs and UHE photons at 100~nG and 316~nG for 1ES 0229+200 assuming the same injection spectra as in figures \ref{fig:var_egmf} and \ref{fig:UHE_photons_with_injection} for the UHECR and neutral channels respectively. 
It is difficult to discriminate the UHECR-induced and UHE-photon-induced cascades only based on their spectra. 
As demonstrated in fig. \ref{fig:var_egmf} lower values of the volume averaged magnetic field strength than $\sim 100$nG in the magnetised region are inconsistent with the spectrum of 1ES 0229+200 if a UHECR origin of the secondary synchrotron emission is assumed. In the UHE photon channel the injection spectrum has been assumed as a representative case of the typical injection expected. For harder injected UHE photon spectra than considered here, a lower magnetic field in the structured region may provide an acceptable fit to the spectrum of 1ES 0229+200. 
Similarly for RGB J0710+591 and 1ES 1218+304 for average magnetic fields below $\sim 100$~nG and $\sim$few~$\times~10$~nG respectively, a UHECR origin would need higher maximum energies. Weaker magnetic fields may also be consistent with a UHE photon origin of the spectrum depending on the details of the injection spectrum.

To summarise, we observe that both channels are a very good fit to the observed gamma-ray spectra although discrimination between the UHE photon and UHECR channels is challenging on the basis of the spectral fit alone for steady gamma-ray sources. The timing properties and the angular extension of the signal contribute towards such a discrimination, for example the observed variability of 1ES 0229+200 and 1ES 1218+304 favour neutral beams as the population responsible for this emission, as discussed in the next section.

\subsection{Time variability}
The main observable differences between UHECR or UHE photon seeding in the magnetised region should be related to the different deflection properties and as a result time delays experienced by the UHE photons and their products in the magnetised region. In the UHE photon channel any deflections will come from the secondary electrons and should be approximately:
\begin{equation} 
\theta_{e} \sim D_{{\rm syn}} / r_{{\rm Lar}}  \sim 3 \times 10^{-4} (E_{e}/10^{19}~{\rm eV})^{-2} (B/10~{\rm nG})^{-1}.
\label{eq:synchrotron_deflection}
\end{equation}
\noindent Typically this is a very small angle, smaller than typical values of $\theta_{{\rm jet}}$ hence the emission from this channel is expected to be beamed. If UHE photons can escape into intergalactic space, the mean free path to $\gamma \gamma$ pair production is $\sim 2$~Mpc, i.e., smaller than the $p \gamma$ energy loss length of UHECRs, so the pair halo/echo signal from UHE neutrals is dominant when the photopion production in the source is efficient (Murase 2012). As a result of the small deflections, the time spread of the signal should also be small. Noting $d \sim \lambda_{\gamma \gamma} \sim 2$~Mpc, \citet{Murase12} obtained $\delta t \sim 2 \theta_{e}^2d/2c \sim 0.3~{\rm yr}~(E_{{\rm syn}}/10^{2.5}~{\rm GeV})^{-2}({\rm min}[d,\lambda_{\gamma \gamma}]/{\rm Mpc})$, where $d$ is the characteristic scale of the magnetised region, as $E_{\rm syn} \propto \gamma_e^2 B$ and $\theta_e \propto \gamma_e^{-2} B^{-1}$.

In comparison the deflections suffered by UHECR protons in the magnetised region are larger, of order $\theta_{p} \sim \sqrt{d~\lambda_{{\rm coh}}}/r_{{\rm Lar}} \sim 0.044~(d/{\rm Mpc})(E/10^{20}~{\rm eV})(\lambda_{{\rm coh}}/d)^{1/2}(B/10~{\rm nG})$. The resulting time spread is also expected to be considerably larger for the UHECR proton channel $\delta t \sim 2 \theta_{p}^2d/2c \sim 1.6 \times 10^{3}~{\rm yr}~(B/10~{\rm nG})(\lambda_{{\rm coh}}/d)(d/{\rm Mpc})^3(E/10^{20}~{\rm eV})^{-2}$. In the case of UHE neutral beams the recent hints of variability of the TeV spectrum of 1ES 0229+200 in $\sim 1$~year timescales can be accommodated. 

For 1ES 1218+304 the model prediction is consistent with the combined GeV-TeV observations of this source. The observed $\sim$day scale variability of this source cannot be explained by our current setup where the UHE photons cascade over an $\sim$Mpc scale structured IGMF region. However, UHE neutral beams could possibly explain the variability of 1ES 1218+304 if the size of the region over which $\gamma \gamma / n \gamma$ interactions occur is significantly smaller (of order kpc) as in the model of \citet{Dermer12} for FSRQs. Detailed work on the variability of these sources will be presented elsewhere.

In the case of blazars a characteristic signature of this channel could be a transient event such as a flare with a duration $\sim 0.1 - 1$~year, whereas the small time spread in the UHE neutral channel implies that we may observe the echo of flaring activities. The two channels may be hard to distinguish between for a given steady gamma-ray source, but if the UHE neutral channel is dominant the resulting emission will almost certainly be more variable.

\section{Discussion, conclusion}\label{section:discussion}

In 50 years of direct searches for the sources of UHECRs we have not made conclusive progress on the subject. The proposed CTA is envisaged to bring about an order of magnitude increase in VHE AGN detections and may thus allow us to make great progress in searches for the secondary emission of UHECRs. Its increased sensitivity by as much as a factor of ten compared to the current generation of IACTs may make it possible to rule out or confirm the existence of the tail expected if the observed VHE emission is due to a UHECR-induced inverse Compton cascade. The absence of such a tail should imply that this emission is due to synchrotron radiation of secondary electrons from UHECRs or of leptonic origin. One can then examine the angular image of the source with the CTA to distinguish between these two possibilities. As shown analytically in \citet{GA05} and modelled numerically in \citet{KAL11} in the secondary electron synchrotron channel one expects a halo to form around the source as a result of the deflection of the primary protons and secondary electrons in the embedding magnetised region. The angular extension of such a halo must be of the order of a fraction of a degree, for sources beyond $\sim 100$~Mpc \citep{Gabici:2006dv} and would therefore appear point-like to the current generation of IACTs, which have a typical point spread function $\sim 0.1^{\circ}$ at $\sim 1$~TeV. Detailed modelling of the halo expected from this channel should make it possible to distinguish it from the halo expected from the leptonic channel \citep[presented in e.g.,][]{Elyiv09,Taylor11} with a CTA type instrument. The HAWC experiment \citep{HAWC2013}, almost fully deployed and currently taking data, will detect transient AGN activity with an unprecedented exposure and should allow significant improvement in our understanding of the time dependence of AGN gamma-ray emission (e.g., \citealp{Imran11}).

As the loss length for protons via Bethe-Heitler pair production is of order $1$~Gpc and through photo-meson production of order $100$~Mpc in the absence of IGMFs one would expect the UHECR cascade signal to dominate over the secondary electron synchrotron signal. This is shown for example in Fig.~\ref{fig:secondary_flux_level} for a source at redshift $z = 0.14$ where we see that the secondary emissions of protons within the magnetised region are $1-2$ orders of magnitude lower than the overall proton losses all the way to the observer. However in the presence of non-negligible IGMFs during propagation the UHECR-induced inverse Compton cascade signal may drop to a lower level than the secondary synchrotron emission. We have shown that, in both cases of UHECR-induced and UHE photon-induced synchrotron cascades, $L_{\rm iso, cr} \sim 10^{46}-10^{47}~{\rm erg}~{\rm s}^{-1}$ is required. Due to the deflection of UHECRs in magnetised regions, similar UHECR luminosities are required in the case of inverse Compton cascades propagating in the intergalactic medium, when UHECR sources are located in filaments but void IGMFs are negligible \citep{Murase:2011cy}. The UHECR-induced inverse Compton cascade signal however will be further suppressed due to the deflection of the charged leptons in the cascade if void IGMFs are non-negligible as we quantify below.

Following equation~(6) of \cite{KAL11}, the gamma-ray flux from a given UHECR source per unit energy interval scales as $f_{\rm 1d}(<B_{\theta}) \,\chi_{e}\, L_{\rm cr}$, where $\chi_{e}L_{{\rm cr}}$ is the luminosity injected in secondary photons and pairs over a distance $d$ by UHECRs and $f_{1d}(<B_{\theta})$ is the one dimensional filling factor of magnetic fields with strength greater than $B_{\theta}$. The fraction of energy $\chi_e$ transferred to pairs and pions is $\sim 0.5$ at $d = 100~{\rm Mpc}$ and grows to $\sim 1$ at $d = 1~{\rm Gpc}$. The magnetic field strength $B_{{\rm \theta}}$ is defined such that the deflection suffered by the low energy electrons at the final stage of the cascade is $\theta$. For $\theta = 1^{\circ}$, $B_{\theta} \sim 2 \times 10^{-14}$~G from equation~(\ref{eq:deflection}) for the low energy electrons in the cascade. The one dimensional filling factor $f_{\rm 1d}$ is the most uncertain quantity in this expression. As discussed in \cite{KAL11}, it is expected that $f_{\rm 1d} \sim f_{\rm3d}$, where $f_{\rm3d}$ is the three dimensional filling factor. Several of the most sophisticated large scale magnetic field models find that a large fraction of the universe is filled with magnetic fields where $B \gg 10^{-14}$~G. It is clear that if this is the case the UHECR-induced inverse Compton cascade signal will be isotropised and hence drop below experimental sensitivity (see e.g., Fig.~5 of \citealp{Ahlers11}, where this effect of homogeneous magnetic fields of strength $B=10^{-14}$~G is illustrated for the case of 1ES 0229+200). If the average magnetic field strength in voids is below $10^{-14}$~G as concluded e.g., in the work of \citet{Donnert09} the expected flux level will depend on the number of magnetised structures crossed. Crossing a galaxy or cluster of galaxies would isotropise the cascade but is unlikely as these occupy a small fraction of the volume in the universe. Crossing a filament is much more likely but with a less dramatic effect on the cascade emission. Assuming that $\sim 10\%$ of the volume is filled with filaments of diameter $\sim$Mpc, crossing the latter would lower the cascade signal level by an order of magnitude. As the UHECR luminosity requirements are already very high, it remains largely dependent on the magnetisation properties of the IGMF whether this channel is ultimately detectable. 

Whether UHECR acceleration is possible in a given source intricately depends on the characteristics of the accelerator. Theoretically, the magnetic luminosity of the relativistic outflow of the source must satisfy $L_B\gtrsim 10^{47.2}\,Z^{-2}E_{20}^2\Gamma_{1}^2~$erg~s$^{-1}$ \citep{Farrar09,LW09,Murase:2011cy}, where $E = 10^{20}\,{\rm eV}/E_{20}$ is the cosmic ray energy and $\Gamma = 10\,\Gamma_1$ is the outflow Lorentz factor. This limit places a stringent constraint on the candidate sources, as these are luminosities typically reached only by FRII-type galaxies, which are the most powerful and rare sources. In this work we have assumed that the maximum acceleration energy in the blazars studied is $E_{p}^{\rm max} \sim 10^{20.5}-10^{21}$~eV. Such energies cannot be achieved by BL Lacs within the SSC model \citep{Murase:2011cy}, which is a drawback of the synchrotron pair halo/echo scenario for extreme TeV blazars. However as noted in \citet{Tavecchio14} the blazar jet parameters obtained in the SSC model for classical BL Lacs can hardly be applicable to the extreme TeV blazars we have studied here. In addition to the conventional shock acceleration mechanism, a number of acceleration mechanisms have been proposed that may allow protons in blazar jets to achieve energies $\sim 10^{20}$~eV. These include the shear acceleration mechanism \citep{RD04} and magnetic reconnection (e.g., \citealp{Giannios10}). As mentioned in the introduction other models exist in which one abandons the SSC interpretation altogether and considers a highly magnetised blazar jet $B \sim 10-100$~G \citep{2000NewA....5..377A, Muecke2003}. In such purely hadronic models proton energies of $10^{20}$~eV can be obtained but in general in these conditions the $p \gamma$ process is inefficient compared to proton synchrotron which is the dominant process responsible for the observed gamma-ray emission. 

Despite the caveat stated above regarding the maximum proton energy, the synchrotron signal we studied in this work is one of the interesting signals and it can coexist with other emission components. Proton acceleration in blazar jets remains highly uncertain, and our assumption for the maximum proton acceleration energy may be accommodated in some non-typical blazars such as extreme TeV blazars. Such high maximum proton acceleration energies in jets have also been discussed in relation to a possible excess of UHECRs in the direction of Cen A (see e.g., \citealp{RA09}). In the near future, detailed multi-wavelength observations of blazar jets will help to clarify the picture. \\
\indent Throughout we have considered sources that emit an isotropic cosmic ray luminosity $L_{{\rm cr, iso}} \sim 10^{47}~{\rm erg}~{\rm s}^{-1}$ above $10^{18}$~eV. Assuming a beaming factor of 100, this corresponds to a beaming-corrected luminosity $L_{\rm cr,~j} \sim 10^{45}~{\rm erg}~{\rm s}^{-1}$. Considering the contribution of low energy protons to the total jet power results in $L_{\rm cr,~j}$ higher by a factor of a few for $\alpha=2$, as discussed in section \ref{sec:results}. This is smaller than the Eddington luminosity, so not unreasonable (although the jet power is typically lower than the Eddington luminosity for BL Lacs, as discussed in \citealp{Ghisellini10}). Note that the required cosmic ray luminosity is at the same level as required by UHECR cascade models. On the other hand, the observed Auger UHECR spectrum suggests that in the Earth's GZK horizon the total power in UHECRs per unit volume per year at energy E is $E^2 {\rm d}N/{\rm d}E \simeq 10^{44}~{\rm erg}~{\rm Mpc}^{-1}~{\rm yr}^{-1}$ assuming an isotropic distribution of sources \citep{2009ApJ...690L..14M,BGG06,Waxman95}. Given that from observations the number density of sources locally is consistent with $n_0 \sim 10^{-5}-10^{-4}~{\rm Mpc}^{-3}$ typical local UHECR sources must have $L_{{\rm cr, j}} \sim 10^{42}~{\rm erg}~{\rm s}^{-1}$, therefore the sources discussed here must be rare and powerful and cannot be typical sources of UHECRs. Due to their distance and the energy losses (the Greisen-Zatsepin-Kuzmin or GZK effect, \citealp{G66,ZK66}) their contribution to the observed UHECR spectrum should be limited to $\lesssim 10\%$. In fact, had any such sources been located within the Earth's GZK horizon one would expect a strong excess in the direction of the source in the arrival direction distribution of UHECRs (see e.g., \citealp{Murase:2011cy, Razzaque12}). \\
\indent In conclusion we have studied the synchrotron halo/echo emission induced by UHECRs and UHE photons in the context of blazars embedded in magnetised regions. In particular, we have demonstrated that the synchrotron emission of UHECR secondaries provides a possible alternative explanation to the more conventional leptonic SSC or UHECR-induced inverse Compton cascade scenarios, for the GeV-TeV spectra of some extreme TeV blazars. As long as UHECR accelerators are located in structured regions with magnetic fields of $\sim10^{-7}$~G, this channel is guaranteed and the flux at the peak energy is insensitive to variations in the overall IGMF strength, which is appealing in view of the large uncertainties on void IGMFs. We also showed that the variability of blazar gamma-ray emission can be accommodated by the synchrotron emission of secondary products of UHE neutrals if these are produced inside accelerators of UHECRs and they are able to escape. The GeV-TeV spectrum of 1ES 0229+200 is consistent with this possible interpretation. The signal, and the model fit to the blazar data in this model depend only on the magnetic field strength in the vicinity of the source and can dominate over other emission components unless the strength of IGMFs in voids is negligibly small. A large fraction of blazars reside in filaments and clusters of large scale structure (e.g., \citealp{Lietzen11} and references therein) and hence this is an almost guaranteed signature as long as the required energy output in hadrons can be met by the accelerator. Finally, we conclude by noting that, the synchrotron cascade signal from magnetised regions studied here and the inverse Compton cascade signal from voids are not mutually exclusive.

\section{Acknowledgement}
We thank S. Ando, O. Lahav, D. Semikoz, H. Takami, A. Taylor and D. Waters for interesting discussions. FO was supported by the Institut Lagrange de Paris (ILP) as a visitor at the Institut d'Astrophysique de Paris. KK thanks the Department of Physics and Astronomy of the University College London for its kind hospitality during this work. KK acknowledges financial support from PNHE and ILP. The work of KM is supported by NASA through Hubble Fellowship, Grant No. 51310.01 awarded by the STScI, which is operated by the Association of Universities for Research in Astronomy, Inc., for NASA, under Contract No. NAS 5- 26555.

\bibliographystyle{aa} 
\bibliography{OMK14} 

\end{document}